\documentclass[twocolumn,pra,aps]{revtex4}
\usepackage{graphicx}

\def\duzomniejsze{<\kern-.7mm<}
\def\duzowieksze{>\kern-.7mm>}

\def\textbf#1{{\bf #1}}
\def\beq{\begin{equation}}
\def\eeq{\end{equation}}
\def\be{\begin{equation}}
\def\ee{\end{equation}}
\def\ben{\begin{eqnarray}}
\def\een{\end{eqnarray}}
\def\beqa{\begin{eqnarray}}
\def\eeqa{\end{eqnarray}}
\def\eea{\end{array}}
\def\bea{\begin{array}}
\newcommand{\bei}{\begin{itemize}}
\newcommand{\eei}{\end{itemize}}
\newcommand{\bee}{\begin{enumerate}}
\newcommand{\eee}{\end{enumerate}}

\def\>{\rangle}
\def\<{\langle}

\begin{document}

\title{Stationary entanglement and nonlocality of two qubits or qutrits collectively
interacting with the thermal environment:
The role of Bell singlet state}

\begin{abstract}
We investigate the stationary entanglement and stationary
nonlocality of two qubits collectively interacting with a common
thermal environment. We assume two qubits are initially in Werner
state or Werner-like state, and find that thermal environment can
make two qubits become stationary nonlocality. The analytical
relations among average thermal photon number of the environment,
entanglement and nonlocality of two qubits are given in details. It
is shown that the fraction of Bell singlet state plays a key role in
the phenomenon that the common thermal reservoir can enhance the
entanglement of two qubits. Moreover, we find that the collective
decay of two qubits in a thermal reservoir at zero-temperature can
generate a stationary maximally entangled mixed state if only the
fraction of Bell singlet state in the initial state is not smaller
than $\frac{2}{3}$. It provides us a feasible way to prepare the
maximally entangled mixed state in various physical systems such as
the trapped ions, quantum dots or Josephson Junctions. For the case
in which two qutrits collectively coupled with a common thermal
reservoir at zero-temperature, we find that the collective decay can
induce the entanglement of two qutrits initially in the maximally
mixed state. The collective decay of two qutrits can also induce
distillable entanglement from the initial conjectured negative
partial
transpose bound entangled states. \\

\end{abstract}
\author{Shang-Bin Li$^{1,2}$}\email{stephenli74@yahoo.com.cn}, \author{Jing-Bo Xu$^{1}$}

\affiliation{1. Zhejiang Institute of Modern Physics and Department
of Physics, Zhejiang University, Hangzhou 310027, People's Republic
of China}
\affiliation{2. Shanghai research center of
Amertron-global,
Zhangjiang High-Tech Park, \\
299 Lane, Bisheng Road, No. 3, Suite 202, Shanghai, 201204, P.R.
China}
\maketitle

\section * {I. INTRODUCTION}

Quantum entanglement plays an important role in quantum
information. It has been recognized as a useful resource in
various quantum information processes \cite{Shor1995}. While
entanglement can be destroyed by the interaction between the
system of interest and its surrounding environment in most
situations, there have been many works showing that the collective
interaction with a common thermal environment can cause the
entanglement of qubits
\cite{Beige2000,Braun2002,Kim2002,Schneider2002,Pleniohue2002,Kraus2004,Clark2003,Duan2003,Zanardi2001,Benatti2003}.
Beige \textit{et al.} have analyzed ways in which entanglement
could be established within a dissipative environments
\cite{Beige2000} and shown that one could even utilize a strong
interaction of the system with its environment to produce
entanglement. Braun has also shown that two qubits with degenerate
energy levels can be entangled via interaction with a common heat
bath \cite{Braun2002}. Schneider and Milburn have studied the
pairwise entanglement in the steady state of the Dicke model and
revealed how the steady state of the ion trap with all ions driven
simultaneously and coupled collectively to a heat bath could
exhibit quantum entanglement \cite{Schneider2002}. Kim \textit{et
al.} have investigated the interaction of the thermal field and a
quantum system composed of two qubits and found that such a
chaotic field with minimal information could entangle qubits that
were prepared initially in a separable state \cite{Kim2002}. Kraus
and Cirac have shown how one could entangle distant atoms by using
squeezed light \cite{Kraus2004}. Clark and Parkins
\cite{Clark2003} have proposed a scheme to controllably entangle
the internal states of two atoms trapped in a high-finesse optical
cavity by employing quantum-reservoir engineering. For generating
multipartite entanglement, Duan and Kimble have proposed an
efficient scheme to engineer multi-atom entanglement by detecting
cavity decay through single-photon detectors \cite{Duan2003}. More
recently, it has been shown that white noise may play a
constructive role in generating the controllable entanglement in
some specific situations \cite{Pleniohue2002,Li2005}.

In this paper, we investigate the system of two qubits or two
qutrits collectively interacting with a common thermal reservoir.
For two-qubit case, we analyze the role of the fraction of Bell
singlet state and the average photon number of thermal reservoir in
the stationary state entanglement and Bell violation. It is shown
that the fraction of Bell singlet state in the initial state is a
key fact determining whether the common thermal reservoir can
enhance the entanglement or Bell violation of two qubits or not. For
two-qutrit case, we find that two qutrits initially in the
conjectured bound entangled Werner state can become distillable due
to the collective decay caused by the common thermal reservoir at
zero-temperature. Even if two qutrits are initially in the maximally
mixed state, they can evolve into a stationary entangled state under
the collective decay. The distinct aspect of collective decay of two
qutrits is that a pure Bell singlet state may be generated from an
initial mixed state.

In the last year, much attention has been paid to the preparation of
the maximally entangled mixed state \cite{Peters2004,Barbieri2004}.
The properties of maximally entangled mixed state have been studied
by many authors \cite{Ishizaka2000,Verstraete2001,Wei2003} Here, we
show that collective decay of two qubits initially in the standard
Werner state in a common thermal reservoir at zero-temperature can
generate a stationary maximally entangled mixed state if only the
fraction of Bell singlet state in the initial state is not smaller
than $\frac{2}{3}$. It is found that stationary state $\rho_1$ of
two qubits initially in the standard Werner state in the common
thermal reservoir builds a bridge across the Werner state with
$r\geq\frac{5}{9}$ and the maximally entangled mixed state if the
temperature of the reservoir can be adiabatically varied from zero
to infinite or vice versa.

This paper is organized as follows: In Sec.II we investigate the
stationary state entanglement of two qubits collectively interacting
with a common thermal reservoir and find that the fraction of the
Bell singlet state in the initial state plays a key role in the
question whether the common thermal reservoir can enhance the
entanglement of two qubits or not. In Sec.III, the Bell violation of
the stationary state of two qubits is investigated and it is shown
that, in certain situation, the common thermal reservoir may drive
two qubits initially satisfying Bell-CHSH inequality into a
stationary state which violates the Bell-CHSH inequality. In Sec.IV,
we investigate the concurrence versus the linear entropy of the
stationary state and find that the common thermal reservoir at
zero-temperature can make two qubits initially in the standard
Werner state become a maximally entangled mixed state if only the
fraction of the Bell singlet state in the initial state is not
smaller than $\frac{2}{3}$. In Sec.V, we turn to consider the case
in which two qutrits collectively interacting with a common thermal
reservoir at zero-temperature and find that the initial maximally
mixed state of two qutrits can become a stationary entangled state.
Furthermore, we show that two qutrits initially in the conjectured
bound entangled Werner state can become free entangled due to the
collective decay caused by the common thermal reservoir at
zero-temperature. In Sec.VI, there are some concluding remarks.

\section * {II. THE STATIONARY ENTANGLEMENT OF TWO QUBITS COLLECTIVELY INTERACTING WITH A THERMAL RESERVOIR}

Up to date, much attention has been paid to the environment-induced
entanglement
\cite{Beige2000,Braun2002,Kim2002,Schneider2002,Pleniohue2002,Kraus2004,Clark2003,Benatti2003}.
Here, we consider such a situation in which two qubits collectively
interacting with a common thermal reservoir. Two qubits are assumed
initially in the Werner state or Werner-like states. Under the
Markovian approximation, the dynamical behavior of two qubits in
this case can be described by the following master equation \beqa
\frac{\partial\hat{\rho}}{\partial{t}}=\frac{(N+1)\gamma}{2}(2\hat{J}_{-}\hat{\rho}\hat{J}_{+}-\hat{J}_{+}\hat{J}_{-}\hat{\rho}-\hat{\rho}\hat{J}_{+}\hat{J}_{-})
\nonumber\\
~~~+\frac{N\gamma}{2}(2\hat{J}_{+}\hat{\rho}\hat{J}_{-}-\hat{J}_{-}\hat{J}_{+}\hat{\rho}-\hat{\rho}\hat{J}_{-}\hat{J}_{+}),
\eeqa where $\gamma$ characterizes the coupling strength between
two qubits and the thermal reservoir. $N$ is the mean phonon
number of the thermal environment. $\hat{J}_{\pm}$ are the
collective atomic operators defined by \beqa
\hat{J}_{\pm}&=&\sum^{2}_{i=1}\hat{\sigma}^{(i)}_{\pm},\nonumber\\
\hat{\sigma}^{(i)}_{+}&=&|1_i\rangle\langle0_i|,~~~\hat{\sigma}^{(i)}_{-}=|0_i\rangle\langle1_i|,
\eeqa where $|1_i\rangle$ and $|0_i\rangle$ are up and down states
of the $i$th qubit, respectively. Recently, the Werner or
Werner-like states \cite{Werner1989,Munro2001,Ghosh2001,Wei2003}
has intrigued many interests for the applications in quantum
information processes. Lee and Kim have discussed the entanglement
teleportation via the Werner states \cite{Lee2000}. Hiroshima and
Ishizaka have studied the entanglement of the so-called Werner
derivative, which is the state transformed by nonlocal
unitary-local or nonlocal-operations from a Werner state
\cite{Hiroshima2000}. Miranowicz has examined the Bell violation
and entanglement of Werner states of two qubits in independent
decay channels \cite{Miranowicz2004}. The experimental preparation
and characterization of the Werner states have also been reported.
An experiment for preparing a Werner state via spontaneous
parametric down-conversion has been put forward \cite{Zhang2002}.
Barbieri \textit{et al.} have presented a novel technique for
generating and characterizing two-photon polarization Werner
states \cite{Barbieri2004}, which is based on the peculiar spatial
characteristics of a high brilliance source of entangled pairs. If
the two qubits are initially prepared in the Werner state or
Werner-like state, how does the external common thermal reservoir
affect their entanglement and nonlocality properties? Here, we
address this question and show that both the stationary state
entanglement and nonlocality heavily depend on the fraction of
Bell singlet state in the initial state and the intensity of the
thermal reservoir. The standard two-qubit Werner state is defined
by \cite{Werner1989} \be
\rho_W=r|\Phi^{-}\rangle\langle\Phi^{-}|+\frac{1-r}{4}I\otimes{I},
\ee where $r\in[0,1]$, and $|\Phi^{-}\rangle$ is the singlet state
of two qubits. $I$ is the identity operator of a single qubit.
Recently, definition (3) is generalized to include the following
states of two qubits \cite{Munro2001,Ghosh2001,Wei2003} \be
\rho^{'}_W=r|\Phi^{+}\rangle\langle{\Phi^{+}}|+\frac{1-r}{4}I\otimes{I},
\ee where
$|\Phi^{\pm}\rangle=\frac{\sqrt{2}}{2}(|10\rangle\pm|01\rangle)$.
Both the Werner state (3) and the Werner-like state (4) are very
important in quantum information. The Werner state (3) is highly
symmetric and $SU(2)\otimes{S}U(2)$ invariant
\cite{Bennett1996,Werner1989}. The mixedness, entanglement and
nonlocality of both the Werner state and the Werner-like state are
uniquely determined by the parameter $r$. In the following, we
consider two different cases. In the case 1, two qubits are
initially in the state ($r\in[0,1]$)
$r|\Phi^{-}\rangle\langle\Phi^{-}|+\frac{1-r}{4}I\otimes{I}$; In
the case 2, two qubits are initially
$r|\Phi^{+}\rangle\langle\Phi^{+}|+\frac{1-r}{4}I\otimes{I}$. The
fractions of the Bell singlet state defined by
${\mathrm{Tr}}(|\Phi^{-}\rangle\langle\Phi^{-}|\rho)$ in both
cases are $f_1(r)=\frac{1+3r}{4}$ and $f_2(r)=\frac{1-r}{4}$
respectively. We will show that the fraction of the Bell singlet
state plays a crucial role in the stationary entanglement and
stationary Bell violation. In the case 1, as the time
$t\rightarrow\infty$, the stationary state of the master equation
(1) can be obtained as follows: \beqa
\rho_1=a_1|11\rangle\langle11|+a_2|10\rangle\langle10|+a_3|01\rangle\langle01|
+a_4|00\rangle\langle00|\nonumber\\
+a_5|10\rangle\langle01|+a^{\ast}_5|01\rangle\langle10|, \eeqa
where \beqa
a_1&=&\frac{(3-3r)N^2}{4L},\nonumber\\
a_2&=&a_3=\frac{r-1+(2+2r)L}{8L},\nonumber\\
a_4&=&1-a_1-a_2-a_3,\nonumber\\
a_5&=&\frac{r-1-4rL}{8L},\nonumber\\
L&=&1+3N(N+1). \eeqa In the case 2, the stationary state of the
master equation (1) can be obtained as follows: \beqa
\rho_2=b_1|11\rangle\langle11|+b_2|10\rangle\langle10|+b_3|01\rangle\langle01|
+b_4|00\rangle\langle00|\nonumber\\
+b_5|10\rangle\langle01|+b^{\ast}_5|01\rangle\langle10|, \eeqa
where \beqa
b_1&=&\frac{(3+r)N^2}{4L},\nonumber\\
b_2&=&b_3=\frac{-r-3+(6-2r)L}{24L},\nonumber\\
b_4&=&1-b_1-b_2-b_3,\nonumber\\
b_5&=&\frac{-r-3+4rL}{24L}. \eeqa In order to quantify the degree of
entanglement, we adopt the concurrence $C$ defined by Wooters
\cite{Woo1998}. The concurrence varies from $C=0$ for an unentangled
state to $C=1$ for a maximally entangled state. For two qubits, in
the "Standard" eigenbasis: $|1\rangle\equiv|11\rangle$,
$|2\rangle\equiv|10\rangle$, $|3\rangle\equiv|01\rangle$,
$|4\rangle\equiv|00\rangle$, the concurrence may be calculated
explicitly from the following: \be
C=\max\{\lambda_1-\lambda_2-\lambda_3-\lambda_4,0\}, \ee where the
$\lambda_{i}$($i=1,2,3,4$) are the square roots of the eigenvalues
\textit{in decreasing order of magnitude} of the "spin-flipped"
density matrix operator
$R=\rho_s(\sigma^{y}\otimes\sigma^{y})\rho^{\ast}_s(\sigma^{y}\otimes\sigma^{y})$,
where the asterisk indicates complex conjugation. It is
straightforward to compute analytically the concurrence $C_1$ and
$C_2$ for the density matrices $\rho_1$ and $\rho_2$, respectively,
\be C_1=\max[0,\frac{1+3r+(18r-6)(N^2+N)}{4(3N^2+3N+1)}], \ee and
\be C_2=\max[0,\frac{1-r-(6+6r)(N+N^2)}{4(3N^2+3N+1)}], \ee which
implies the stationary state $\rho_1$ is entangled for the case with
a fixed $N$ if and only if $r>\frac{6N^2+6N-1}{18N^2+18N+3}$.
Meanwhile, the stationary state $\rho_2$ is entangled if and only if
two inequalities $0\leq{r}<\frac{1-6N-6N^2}{1+6N+6N^2}$ and
$0\leq{N}<\frac{\sqrt{15}-3}{6}$ are simultaneously satisfied. In
Fig.1, the concurrences $C_1$ is plotted as the function of the
parameter $r$ of the initial Werner state and the average phonon
number $N$ of the thermal reservoir. It is shown that, if two qubits
are initially in the standard Werner state, the stationary
entanglement of two qubits increases with the fraction of the Bell
singlet state in their initial state, and decreases with $N$. In
Fig.2, we plot the concurrence $C_2$ as the function of the
parameter $r$ of the initial Werner-like state in Eq.(4) and the
average photon number $N$ of the thermal reservoir. We can see that
the stationary state entanglement decreases both with the increase
of $r$ and $N$, which implies the higher initial entanglement does
not result in the higher stationary state entanglement. The reason
is that the fraction of the Bell singlet state in the Werner-like
state $r|\Phi^{+}\rangle\langle\Phi^{+}|+\frac{1-r}{4}I\otimes{I}$
decreases with $r$. From Fig.1 and Fig.2, we can observe that, in
the case with $N=0$, namely two qubits collectively interact with a
thermal reservoir at zero-temperature, the stationary state is
always entangled if only the initial state of two qubits is not
absolutely symmetric, i.e. the fraction of Bell singlet state in the
initial state is not zero. When $r\leq\frac{1}{3}$, the initial
Werner state or Werner-like state are separable. Surprisingly, the
external thermal reservoir can enhance the entanglement of two
qubits even if two qubits are initially in a separable state. For
example, when $r=0$, two qubits is initially in the maximally mixed
state. However, the collective interaction between two qubits and
the low-temperature thermal reservoir can drive two initial
maximally mixed qubits into a stationary entangled mixed state.
Comparing the stationary state entanglement with the entanglement of
the initial Werner or Werner-like states, we can obtain the
condition that the entanglement can be enhanced by external common
thermal environment. This condition closely relates to the fraction
$F$ of Bell singlet state and the average photon number of the
thermal environment, which can be expressed by the following
inequalities: \be 1>F>\max(\frac{1}{6},\frac{3N^2+3N}{6N^2+6N+1}).
\ee or \be \frac{1}{6}\geq{F}>\frac{9N^2+9N+2}{36N^2+36N+14}. \ee
The increment of entanglement between the stationary state
entanglement and its initial entanglement of the Werner state can be
obtained as follows: \beqa
\Delta{C}=\max[0,C_1-\max(0,\frac{3r-1}{2})]~~~{\mathrm{For~case~1
}},\nonumber\\
\Delta{C}=\max[0,C_2-\max(0,\frac{3r-1}{2})]~~~{\mathrm{For~case~2
}}.\eeqa In Fig.3, we plot the increment of entanglement between
the stationary state entanglement and its initial entanglement of
the Werner state as the function of $N$ and $r$. It is shown that
the thermal reservoir can enhance the entanglement of two qubits
in the range indicated by the inequalities (12) and (13). When the
parameter $r$ of the initial Werner state is larger than
$\frac{1}{3}$, the increment of entanglement $\Delta{C}$ decreases
both with the increase of $r$ and $N$. When $r\leq\frac{1}{3}$,
the initial state is separable, nevertheless, the stationary state
may be entangled if only $r>\frac{6N^2+6N-1}{18N^2+18N+3}$. From
Eq.(10) or Eq.(14), we can get a conclusion that the common
thermal reservoir with any large intensity can enhance the
entanglement of two qubits collectively coupled with the reservoir
if only the fraction $F$ of Bell singlet state in the initial
Werner state is smaller than 1 and not smaller than $\frac{1}{2}$.
In Fig.4, the increment of entanglement between the stationary
state entanglement and its initial entanglement of the Werner-like
state in Eq.(4) is plotted as the function of $N$ and $r$. We can
see that, in this case, the increment of entanglement decreases
both with $r$ and $N$. In the following section, we shall
investigate how a common thermal reservoir affects the Bell
violation of two qubits initially in the Werner state.
\begin{figure}
\centerline{\includegraphics[width=2.5in]{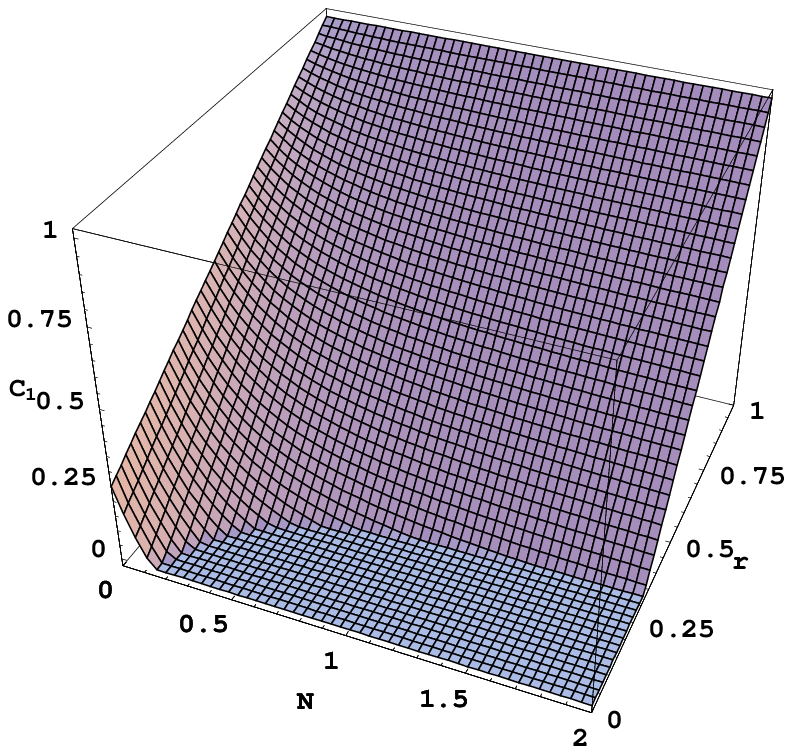}}
\caption{The concurrence $C_1$ of the stationary states $\rho_1$
is plotted as the function of the parameter $r$ of the initial
Werner state
$r|\Phi^{-}\rangle\langle\Phi^{-}|+\frac{1-r}{4}I\otimes{I}$ and
the intensity $N$ of the external common thermal environment.}
\end{figure}
\begin{figure}
\centerline{\includegraphics[width=2.5in]{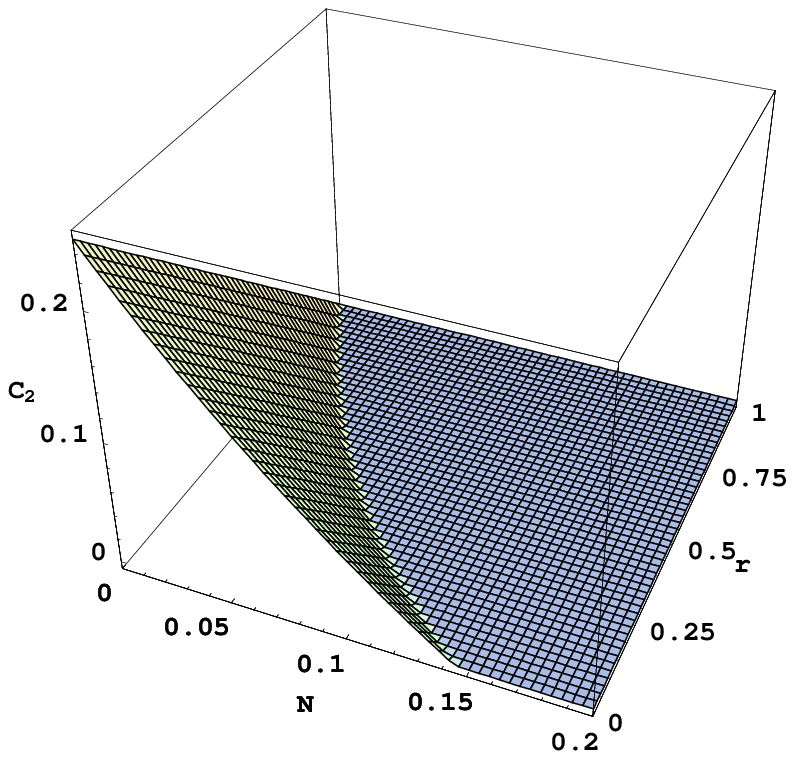}}
\caption{The concurrence $C_2$ of the stationary states $\rho_2$
is plotted as the function of the parameter $r$ of the initial
Werner-like state
$r|\Phi^{+}\rangle\langle\Phi^{+}|+\frac{1-r}{4}I\otimes{I}$ and
the intensity $N$ of the external common thermal environment.}
\end{figure}
\begin{figure}
\centerline{\includegraphics[width=2.5in]{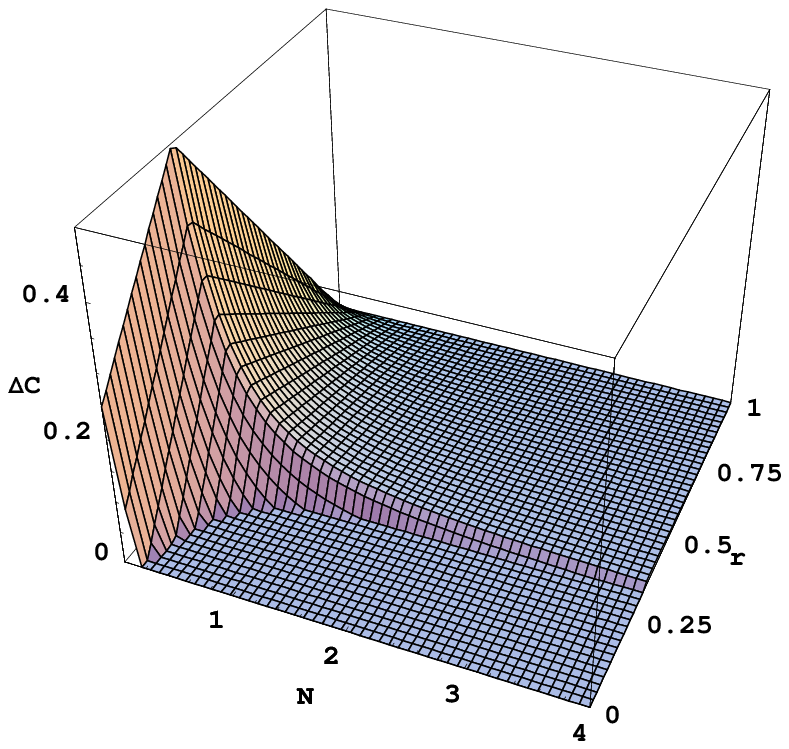}}
\caption{The increment $\Delta{C}$ of entanglement between the
stationary state entanglement and its initial entanglement of the
Werner state
$r|\Phi^{-}\rangle\langle\Phi^{-}|+\frac{1-r}{4}I\otimes{I}$ is
plotted as the function of $N$ and $r$.}
\end{figure}

\begin{figure}
\centerline{\includegraphics[width=2.5in]{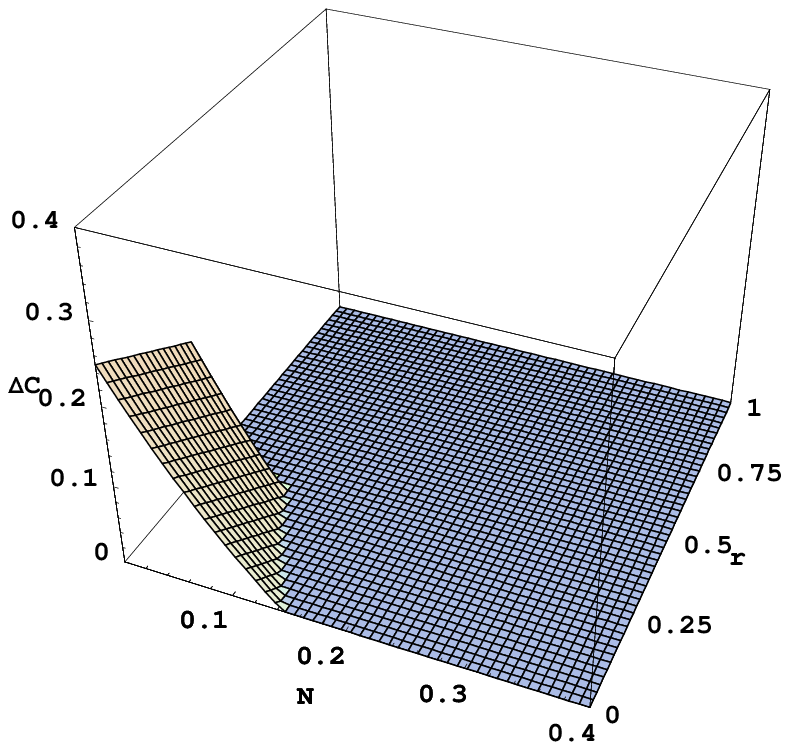}}
\caption{The increment $\Delta{C}$ of entanglement between the
stationary state entanglement and its initial entanglement of the
Werner-like state
$r|\Phi^{+}\rangle\langle\Phi^{+}|+\frac{1-r}{4}I\otimes{I}$ is
plotted as the function of $N$ and $r$.}
\end{figure}

\section * {III. BELL VIOLATION OF TWO QUBITS INTERACTING WITH A COMMON THERMAL ENVIRONMENT}

In this section, we attempt to discuss the nonlocality of two
qubits in their stationary states. The nonlocal property of two
qubits can be characterized by the maximal violation of Bell
inequality. Recently, it has been argued that entanglement and
nonlocality of two qubits are different resources
\cite{Brunner2005}. We also find that the
stochastic-resonance-like behavior of entanglement can not be
observed in the Bell violation of two qubits during the evolution
\cite{Li2005}. The concurrence, one of the good entanglement
measures, is not monotonic function of the maximal violation of
Bell inequality for some entangled mixed states of two qubits
\cite{Vers2002}. So it is interesting to investigate how the
collective decay of two qubits can affect their maximal violation
of Bell inequality. The most commonly discussed Bell inequality is
the CHSH inequality \cite{Bell1965,CHSH}. The CHSH operator reads
\be
\hat{B}=\vec{a}\cdot\vec{\sigma}\otimes(\vec{b}+\vec{b^{\prime}})\cdot\vec{\sigma}
+\vec{a^{\prime}}\cdot\vec{\sigma}\otimes(\vec{b}-\vec{b^{\prime}})\cdot\vec{\sigma},
\ee where $\vec{a},\vec{a^{\prime}},\vec{b},\vec{b^{\prime}}$ are
unit vectors. In the above notation, the Bell inequality reads \be
|\langle\hat{B}\rangle|\leq2. \ee The maximal amount of Bell
violation of a state $\rho$ is given by \cite{Horo1995} \be
{\mathcal{B}}=2\sqrt{\lambda+\tilde{\lambda}}, \ee where $\lambda$
and $\tilde{\lambda}$ are the two largest eigenvalues of
$T^{\dagger}_{\rho}T_{\rho}$. The matrix $T_{\rho}$ is determined
completely by the correlation functions being a $3\times3$ matrix
whose elements are
$(T_{\rho})_{nm}={\mathrm{Tr}}(\rho\sigma_{n}\otimes\sigma_{m})$.
Here, $\sigma_1\equiv\sigma_x$, $\sigma_2\equiv\sigma_y$, and
$\sigma_3\equiv\sigma_z$ denote the usual Pauli matrices. We call
the quantity $\mathcal{B}$ the maximal violation measure, which
indicates the Bell violation when ${\mathcal{B}}>2$ and the
maximal violation when ${\mathcal{B}}=2\sqrt{2}$. In what follows,
we focus our attention on the role of the common thermal reservoir
on the Bell violation of two qubits. If two qubits are initially
in the Werner or Werner-like states, we find that the Bell
violation of stationary state of two qubits heavily depends on the
fraction of the Bell singlet state in the initial state.

\begin{figure}
\centerline{\includegraphics[width=2.5in]{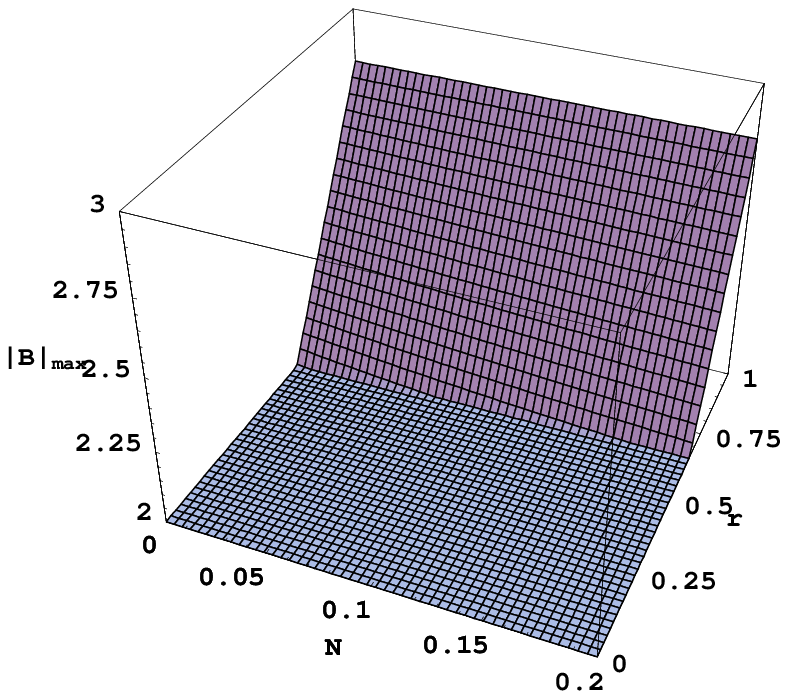}}
\caption{Maximal Bell violation of the stationary state $\rho_1$
is plotted as the function of the parameter $r$ of the initial
Werner state
$r|\Phi^{-}\rangle\langle\Phi^{-}|+\frac{1-r}{4}I\otimes{I}$ and
the intensity $N$ of the external common thermal environment.}
\end{figure}
\begin{figure}
\centerline{\includegraphics[width=2.5in]{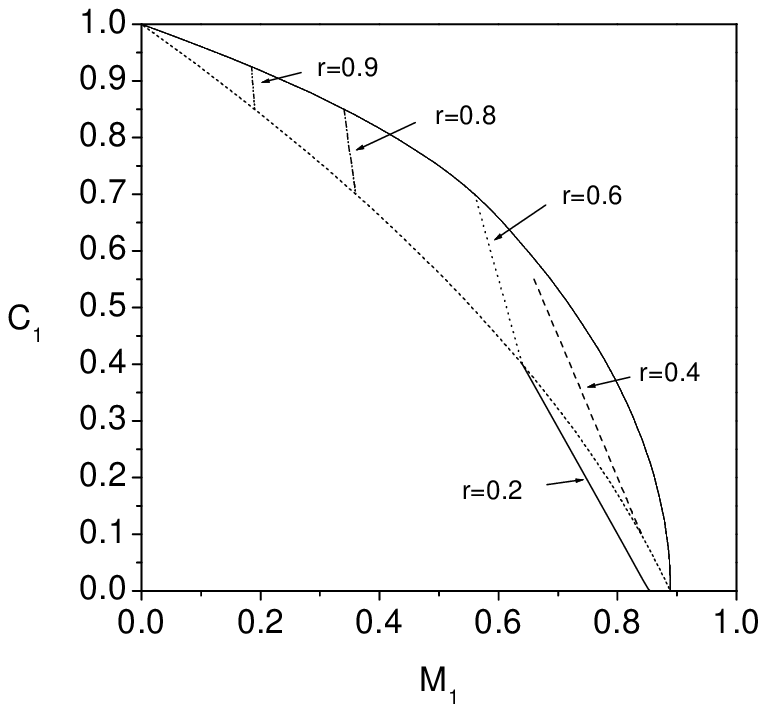}}
\caption{Concurrence versus mixedness for the stationary states
$\rho_1$ is depicted for any possible values of $N$ and different
values of $r$. The short dash line and the short dot line
represent the Werner state and the maximally entangled mixed state
(the frontier of the concurrence versus the linear entropy),
respectively.}
\end{figure}

For the density operator $\rho_1$ in Eqs.(5-6) and $\rho_2$ in
Eqs.(7-8) characterizing the two stationary states of two qubits
governed by the master equation (1) corresponding to two kinds of
different initial states, the maximal Bell violation $|B_1|_{max}$
and $|B_2|_{max}$ can be written as follows \be
|B_1|_{max}=2\sqrt{4|a_5|^2+\max[4|a_5|^2,(1-4a_2)^2]} \ee and \be
|B_2|_{max}=2\sqrt{4|b_5|^2+\max[4|b_5|^2,(1-4b_2)^2]} \ee The
sufficient and necessary condition for $|B_1|_{max}>2$ can be
derived as follows: \be
r>\frac{2\sqrt{2}-1+6\sqrt{2}N(N+1)}{3+12N(N+1)}, \ee and
$|B_2|_{max}$ can be easily verified that it can not be larger
than $2$. Since the initial standard Werner state can not violate
any Bell-CHSH inequality when $r\leq\frac{\sqrt{2}}{2}$, it is
interesting that the corresponding stationary state may achieve
the Bell violation even if
$\frac{\sqrt{2}}{2}\geq{r}>\frac{2\sqrt{2}-1+6\sqrt{2}N(N+1)}{3+12N(N+1)}$,
which means the common thermal reservoir can induce the stationary
Bell violation of two qubits. This may be important for the
experimental verification of Bell violation in the quantum dots in
which the collective decay may be caused by the common thermal
phonon background. In Fig.5, we plot the maximal value of the Bell
violation of the stationary state as the function of $r$ and $N$
for the case in which two qubits are initially in the standard
Werner state. It is shown that Bell violation of the stationary
state decreases with the decrease of the parameter $r$. The Bell
violation also decreases with $N$. However, if the initial
standard Werner state is very pure, the Bell violation of its
stationary state is robust against the collective decay.

\begin{figure}
\centerline{\includegraphics[width=2.5in]{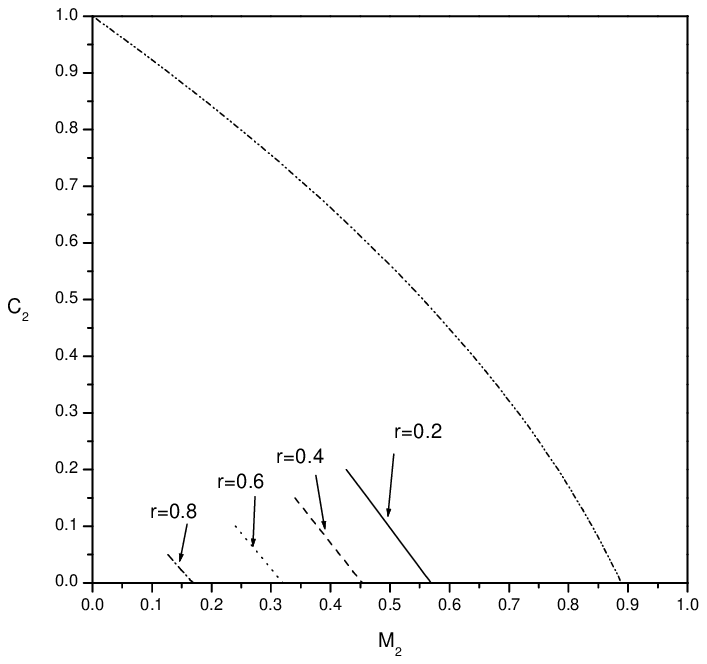}}
\caption{Concurrence versus mixedness for the stationary states
$\rho_2$ is depicted for any possible values of $N$ and different
values of $r$. The dash dot dot line represents the Werner state.}
\end{figure}

\section * {IV. CONCURRENCE VERSUS MIXEDNESS OF TWO QUBITS INTERACTING WITH A COMMON THERMAL ENVIRONMENT}

In this section, we pay our attention to investigate how the
common thermal reservoir affects the relation between the
concurrence and the mixedness of the stationary state. We find the
common thermal reservoir can drive two qubits to exceed the curve
of the initial standard Werner state in the figure labelled by the
concurrence and the linear entropy if only the fraction of Bell
singlet state is not smaller than a threshold value. Ordinarily,
the mixedness of a state can be characterized by the linear
entropy which is defined by
$M=\frac{4}{3}(1-{\mathrm{Tr}}\rho^2)$. For the stationary states
$\rho_1$ and $\rho_2$ in Eqs.(5-6) and Eqs.(7-8) respectively, the
mixedness can be calculated as follows: \beqa
M_1(\rho_1)=\frac{4}{3}(1-\sum^4_{i=1}a^2_i-2a^2_5),\nonumber\\
M_2(\rho_2)=\frac{4}{3}(1-\sum^4_{i=1}b^2_i-2b^2_5). \eeqa In
Fig.6, we display the concurrence and the mixedness of the
stationary state $\rho_1$ of two qubits which is initially in the
standard Werner state. From Fig.6, it can be observed that in the
situations with $r\geq0.4$, the corresponding stationary state can
go beyond the curve of the concurrence and linear entropy of the
original Werner state. When the intensity of the common thermal
reservoir decreases, the concurrence of the corresponding
stationary state $\rho_1$ increases and its mixedness
characterized by the linear entropy decreases. This implies that
the common thermal reservoir not only can enhance the concurrence
of the mixed state initially with large fraction of Bell singlet
state but also can decrease the mixedness. This counterintuitive
phenomenon may be helpful for the entanglement purification or
distillation.

From Eqs.(5-6), we can immediately know that the stationary state is
as the same as the initial Werner state in the case of infinite high
temperature thermal reservoir, i.e. $N\rightarrow\infty$. In the
case with $N=0$, i.e. a thermal reservoir at zero-temperature, the
corresponding stationary states $\rho_1$ with $r\geq\frac{5}{9}$
become the maximally entangled mixed state, i.e. the frontier of the
concurrence versus the linear entropy of two qubits \cite{Wei2003}.
So we can achieve a conclusion that the common thermal reservoir at
zero-temperature can make two qubits initially in the standard
Werner state become a maximally entangled mixed state if only the
fraction of the Bell singlet state in the initial state is not
smaller than $\frac{2}{3}$. It provides us a feasible way to prepare
the maximally entangled mixed state in various physical systems such
as the trapped ions, quantum dots or Josephson Junctions. In these
systems, the collective decay has been extensively studied both
theoretically and experimentally. One may see that stationary state
$\rho_1$ of two qubits in the common thermal reservoir builds a
bridge across the Werner state with $r\geq\frac{5}{9}$ and the
maximally entangled mixed state if the temperature of the reservoir
can be adiabatically varied. We conjecture that this property is
closely related to the fact that the frontier of the concurrence
versus the linear entropy of two qubits contains two different
branches \cite{Wei2003}. Interestingly, If we adopt negativity to
measure the entanglement of two qubits, the Werner state becomes the
frontier in the sense that these states have the maximal negativity
for a given linear entropy or Von-Neumann entropy \cite{Wei2003}. So
roughly speaking, the stationary states $\rho_1$ with $r\geq5/9$ of
two qubits in the common thermal reservoir in two extreme
situations, i.e. the zero temperature and the infinite high
temperature, become part of the frontier of the concurrence versus
linear entropy and the whole frontier of the negativity versus
linear entropy, respectively. Therefore, it is very necessary to
study the relation among the negativity of the stationary state, the
intensity of the common reservoir and the fraction of Bell singlet
state.

The negativity for a bipartite state $\rho$ is defined as \be
{\mathcal{N}}(\rho)=2\sum_i|\mu_i|, \ee where $\mu_i$ is the
negative eigenvalues of partial transpose $\rho^{\Gamma}$ of the
density matrix $\rho$. We can easily obtain the negativity
${\mathcal{N}}_1$ and ${\mathcal{N}}_2$ of the stationary state
$\rho_1$ in Eqs.(5-6) and $\rho_2$ in Eqs.(7-8) respectively as
follows: \beqa
{\mathcal{N}}_1&=&\frac{1}{2}|a_1+a_4-\sqrt{(a_1-a_4)^2+4|a_5|^2}|\nonumber\\
&&-\frac{1}{2}(a_1+a_4-\sqrt{(a_1-a_4)^2+4|a_5|^2}), \eeqa and
\beqa
{\mathcal{N}}_2&=&\frac{1}{2}|b_1+b_4-\sqrt{(b_1-b_4)^2+4|b_5|^2}|\nonumber\\
&&-\frac{1}{2}(b_1+b_4-\sqrt{(b_1-b_4)^2+4|b_5|^2}), \eeqa We find
that the negativity ${\mathcal{N}}_1$ decreases with $N$ and
increases with $r$, and the negativity ${\mathcal{N}}_2$ decreases
with $N$ and $r$. In Fig.7, we display the concurrence and the
mixedness of the stationary state $\rho_2$ of two qubits. It is
shown that, the entanglement of the stationary state for a given
value of mixedness is much smaller than the entanglement of the
Werner state with the same value of mixedness.

\section * {V. THE COLLECTIVE DECAY OF TWO QUTRITS}

In this section, we turn to consider the collective decay of two
qutrits in the common thermal reservoir at zero-temperature. Under
the Markovian approximation, the collective decay of two qutrits can
be described by the following master equation: \be
\frac{\partial\rho}{\partial{t}}=\gamma(2L_{-}\rho{L}_{+}-L_{+}L_{-}\rho-\rho{L}_{+}L_{-}),\ee
where $L_{\pm}\equiv\sum^{2}_{i=1}J^{(i)}_{\pm}$, and $J^{(i)}_{-}$
($J^{(i)}_{+}$) is the qutrit down (up) operator of the $i$th
qutrit. The representation of $J_{\pm}$ in the space spanned by the
three orthogonal vector $\{|1\rangle,|2\rangle,|3\rangle\}$ of a
qutrit can be written as \beqa
J_{-}=\sqrt{2}|1\rangle\langle2|+\sqrt{2}|2\rangle\langle3|\nonumber\\
J_{+}=\sqrt{2}|2\rangle\langle1|+\sqrt{2}|3\rangle\langle2|.\eeqa
If we assume that two qutrits are initially in the maximally mixed
state, i.e.
$\rho_0=\frac{1}{9}\sum^{3}_{i,j=1}|i,j\rangle\langle{i,j}|$, then
the corresponding stationary state of the master equation (24) can
be obtained as \beqa
\rho_s=\frac{5}{9}|1,1\rangle\langle1,1|+\frac{1}{6}(|1,2\rangle-|2,1\rangle)(\langle1,2|-\langle2,1|)\nonumber\\
+\frac{1}{27}(|3,1\rangle+|1,3\rangle-|2,2\rangle)(\langle3,1|+\langle1,3|-\langle2,2|).\eeqa
It is easy to verify that the $\rho_s$ is an entangled state of
two qutrits and its negativity defined by Eq.(22) can be
calculated as $\frac{\sqrt{97}-8}{27}$. In what follows, we
further show that two qutrits initially in some conjectured
negative partial transpose bound entangled states can become
distillable. In Ref.\cite{DiVincenzo2000,Dur2000}, the authors
presented the following conjecture: Given is the class of Werner
state in ${\mathcal{H}}_3\otimes{\mathcal{H}}_3$ \be
\rho_W(\eta)=\frac{1}{8\eta-1}(\eta{I_3}\otimes{I_3}-\frac{\eta+1}{3}\sum^{3}_{i,j=1}|i,j\rangle\langle{j,i}|)
\ee where $I_3$ is the identity operator in ${\mathcal{H}}_3$. The
state $\lim_{\eta\rightarrow\infty}\rho_W(\eta)$ is separable and
for any finite $\eta\geq\frac{1}{2}$ $\rho_W(\eta)$ is entangled
and violates the Peres-Horodecki criterion
\cite{Peres1996,Horodecki1996}. It has been shown that there is
convincing evidence in support of the conjecture that for all
$\eta\geq2$ the state $\rho_W(\eta)$ is undistillable. We will
show that two qutrits initially in $\rho_W(\eta)$ can become a
stationary free entangled state.

Substituting $\rho_W(\eta)$ into Eq.(25), and the corresponding
stationary state can be easily obtained \beqa
\rho^{(s)}_W(\eta)=\frac{1}{24\eta-3}[(10\eta-5)|1,1\rangle\langle1,1|+(2\eta-1)|S_1\rangle\langle{S}_1|\nonumber\\
+(12\eta+3)|A_1\rangle\langle{A}_1|],\eeqa where
$|S_1\rangle=\frac{\sqrt{3}}{3}(|3,1\rangle+|1,3\rangle-|2,2\rangle)$,
and $|A_1\rangle=\frac{\sqrt{2}}{2}(|1,2\rangle-|2,1\rangle)$. The
negativity of the stationary state $\rho^{(s)}_W(\eta)$ can be
calculated as \be
N=2(\sqrt{\zeta^2_3+4\zeta^2_2}-\zeta_3)+2|\kappa|,\ee where
$\kappa$ is the negative root of the polynomial \be
x^3-(\zeta_1+\zeta_2)x^2+(\zeta_1\zeta_2-\zeta^2_2-\zeta^2_3)x+\zeta^3_2=0,
\ee and \beqa \zeta_1=\frac{10\eta-5}{24\eta-3},\nonumber\\
\zeta_2=\frac{2\eta-1}{72\eta-9},\nonumber\\
\zeta_3=\frac{4\eta+1}{16\eta-2}. \eeqa
\begin{figure}
\centerline{\includegraphics[width=2.5in]{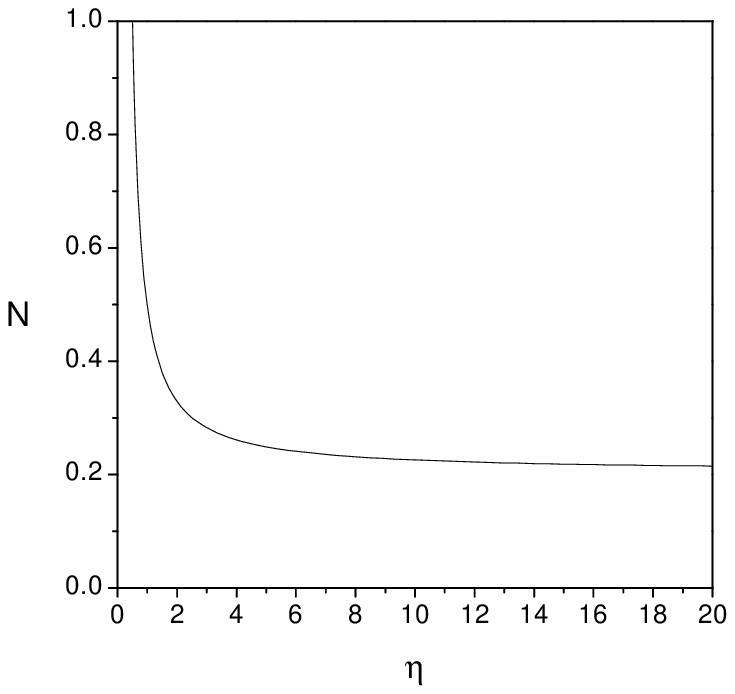}}
\caption{The negativity of the stationary state
$\rho^{(s)}_W(\eta)$ is plotted as the function of the parameter
$\eta$.}
\end{figure}
In Fig.8, we plot the negativity $N$ of the stationary state
$\rho^{(s)}_W(\eta)$ as the function of $\eta$. It is shown that
the negativity decreases with $\eta$ and eventually converges to
about 0.21. In the case with $\eta=\frac{1}{2}$, we can see that
the stationary state $\rho^{(s)}_W(\frac{1}{2})$ is a maximally
entangled state $|A_1\rangle\langle{A}_1|$ in the space spanned by
$\{|1,1\rangle,|1,2\rangle,|2,1\rangle,|2,2\rangle\}$. This
implies that the common zero-temperature thermal reservoir plays a
similar role with the entanglement purifier in this case. In the
limit $\eta\rightarrow\infty$, $\rho^{(s)}_W(\infty)$ is also
entangled. We will demonstrate that $\rho^{(s)}_W(\eta)$
($\eta\in[2,\infty)$) is distillable even though the initial state
$\rho_W(\eta)$ is conjectured to be undistillable in this region.
By applying the local projection
$\Pi_{1,2}\otimes\Pi_{1,2}\equiv(|1\rangle\langle1|+|2\rangle\langle2|)\otimes(|1\rangle\langle1|+|2\rangle\langle2|)$
on $\rho^{(s)}_W(\eta)$, we can immediately obtain the resulted
density operator \beqa
\rho^{(r)}_W(\eta)=\frac{\Pi_{1,2}\otimes\Pi_{1,2}
\rho^{(s)}_W(\eta)\Pi_{1,2}\otimes\Pi_{1,2}}{{\mathrm{Tr}}(\Pi_{1,2}\otimes\Pi_{1,2}
\rho^{(s)}_W(\eta)\Pi_{1,2}\otimes\Pi_{1,2})}\nonumber\\
=\frac{2\eta-1}{68\eta-7}|2,2\rangle\langle2,2|+\frac{30\eta-15}{68\eta-7}|1,1\rangle\langle1,1|\nonumber\\
+\frac{36\eta+9}{68\eta-7}|A_1\rangle\langle{A}_1|,\eeqa where
${\mathrm{Tr}}(\Pi_{1,2}\otimes\Pi_{1,2}
\rho^{(s)}_W(\eta)\Pi_{1,2}\otimes\Pi_{1,2})=\frac{68\eta-7}{72\eta-9}$
is the probability of obtaining the state $\rho^{(r)}_W(\eta)$. It
is easy to verify that the partial transpose of
$\rho^{(r)}_W(\eta)$ is not positive for any values of
$\eta\in[\frac{1}{2},\infty)$. So $\rho^{(r)}_W(\eta)$ is always
distillable for any values of $\eta\in[\frac{1}{2},\infty)$
according to the fact that the nonpositivity of a partial
transpose is necessary and sufficient for distillability of
$2\times2$ systems \cite{Horodecki1997}. Since the state
$\rho^{(r)}_W(\eta)$ comes from the local projection of the state
$\rho^{(s)}_W(\eta)$, the distillability of $\rho^{(r)}_W(\eta)$
indicates that $\rho^{(s)}_W(\eta)$ is also distillable in the
region $\eta\in[\frac{1}{2},\infty)$. This kind of
environment-assisted distillation may be very important for
quantum information processes based on qutrits. It is possible to
generalize the above results to two qudits with the higher
dimension or two spins $\frac{d-1}{2}$ with $d>3$. Those results
will be discussed elsewhere.

\section * {VI. CONCLUSION}

In this paper, we have investigated the systems of two qubits or
qutrits collectively interacting with a common thermal reservoir.
It is shown that the fraction of Bell singlet state in the initial
state is a key fact determining whether the collective decay can
enhance the stationary state entanglement or Bell violation of two
qubits or not. We have also found that collective decay of two
qubits or two qutrits can induce stationary entanglement from
their initial maximally mixed state. The detailed analytical
relations among average thermal phonon number of the thermal
reservoir, entanglement and Bell violation of two qubits have been
obtained. The common thermal reservoir with any large intensity
can enhance the entanglement of two qubits initially in a mixed
Werner state collectively coupled with the reservoir if only the
fraction $F$ of Bell singlet state in the initial state is not
smaller than $\frac{1}{2}$. If the fraction $F$ of Bell singlet
state in the initial Werner state is not smaller than
$\frac{2}{3}$, two qubits in a common zero-temperature thermal
reservoir can evolve into a stationary maximally entangled mixed
state as the time $t\rightarrow\infty$. The corresponding
stationary states of two qubits initially in the standard Werner
state with $r\geq5/9$ in the common thermal reservoirs with two
extreme situations, i.e. the zero temperature and the infinite
high temperature, become part of the frontier of the concurrence
versus linear entropy and the whole frontier of the negativity
versus linear entropy, respectively.

For two-qutrit case, we have found that two qutrits initially in
the conjectured bound entangled Werner state can become
distillable due to the collective decay caused by the common
zero-temperature thermal reservoir. In addition, we obtained a
more striking result that a pure stationary Bell singlet state in
the $2\times2$ subspace of two qutrits may be generated from an
initial mixed state in such a decoherence process. This kind of
environment-induced superselection \cite{Zurek2003} may provide us
a very useful quantum channel which is very desirable in quantum
communication or quantum computation.

\section*{ACKNOWLEDGMENT}
This project was supported by the National Natural Science
Foundation of China (Project NO. 10174066).

\bibliographystyle{apsrev}
\bibliography{refmich,refjono}

\end{document}